\documentstyle[aps,prl,preprint,floats,epsfig]{revtex}

\def\mtiny{\vrule width 0pt}
\def\mrm#1{\mathrm{#1}}
\def\DZ{\relax\ifmmode{D^0}\else{$\mrm{D}^{\mrm{0}}$}\fi}
\def\DONE{\relax\ifmmode{D_1}\else{$\mrm{D}_{\mrm{1}}$}\fi}
\def\DTWO{\relax\ifmmode{D_2}\else{$\mrm{D}_{\mrm{2}}$}\fi}
\def\KZ{\relax\ifmmode{K^0}\else{$\mrm{K}^{\mrm{0}}$}\fi}
\def\KSHO{\relax\ifmmode{K_{\rm S}}\else{$\mrm{K}_{\mrm{S}}$}\fi}
\def\KLON{\relax\ifmmode{K_{\rm L}}\else{$\mrm{K}_{\mrm{L}}$}\fi}
\def\BZ{\relax\ifmmode{B^0_d}\else{$\mrm{B}^{\mrm{0}_d}$}\fi}\def\BZp{\relax\ifmmode{B^0}\else{$\mrm{B}^{\mrm{0}}$}\fi}
\def\BZS{\relax\ifmmode{B^0_s}\else{$\mrm{B}^{\mrm{0}_s}$}\fi}
\def\DZS{\relax\ifmmode{D^{*+}}\else{$\mrm{D}^{\mrm{*+}}$}\fi}
\def\DZB{\relax\ifmmode{\overline{D}\mtiny^0}
        \else{$\overline{\mrm{D}}\mtiny^{\mrm{0}}$}\fi}\def\KZB{\relax\ifmmode{\overline{K}\mtiny^0}
        \else{$\overline{\mrm{K}}\mtiny^{\mrm{0}}$}\fi}
\def\BZB{\relax\ifmmode{\overline{B}\mtiny^0_d}
        \else{$\overline{\mrm{B}}\mtiny^{\mrm{0}_d}$}\fi}
\def\BZBp{\relax\ifmmode{\overline{B}\mtiny^0}
        \else{$\overline{\mrm{B}}\mtiny^{\mrm{0}}$}\fi}
\def\BZBS{\relax\ifmmode{\overline{B}\mtiny^0_s}
        \else{$\overline{\mrm{B}}\mtiny^{\mrm{0}_s}$}\fi}
\def\DZC{\relax\ifmmode{\overline{D}\mtiny^0}
        \else{$\overline{\mrm{D}}\mtiny^{\mrm{0}}$}\fi}

\begin{document}


\tighten

\preprint{\tighten\vbox{\hbox{\hfil CLNS 99/1659}
                        \hbox{\hfil CLEO 99-23}
}}
\title{Search for $\boldmath\DZ\!-\!\DZB$ Mixing}

\date{December 31, 1999}
\maketitle

\begin{abstract}
We report on a search for $\DZ\!-\!\DZB$ mixing made by
studying the `\hbox{wrong-sign}' process
$\DZ\!\to\!K^+\pi^-$.
The data come from an
integrated luminosity 
of 9.0~fb$^{-1}$ of $e^+e^-$
collisions at $\sqrt{s}\approx10\,$GeV
recorded with the CLEO II.V detector.
We measure the time integrated rate of the `\hbox{wrong-sign}' process
$\DZ\!\to\!K^+\pi^-$ relative to that of the Cabibbo-favored
process $\DZB\!\to\!K^+\pi^-$ to be
$R=(0.332^{+0.063}_{-0.065}\pm0.040)\%$.
We study $\DZ\!\to\!K^+\pi^-$ as a function
of decay time to distinguish
direct doubly Cabibbo-suppressed decay from
$\DZ\!-\!\DZB$ mixing.
The amplitudes that describe
$\DZ\!-\!\DZB$ mixing, $x^\prime$ and $y^\prime$, are consistent
with zero. At the 95\% C.L. and without assumptions 
concerning charge-parity (CP) violating parameters,
we find $(1/2)x^{\prime2}<0.041\%$ and $-5.8\%\!<\!y^\prime\!<\!1.0\%$.
\end{abstract}
\pacs{Pacs number 14.40.Lb,13.25.Ft}
\newpage

{
\renewcommand{\thefootnote}{\fnsymbol{footnote}}

\begin{center}
R.~Godang,$^{1}$ K.~Kinoshita,$^{1,}$%
\footnote{Permanent address: University of Cincinnati, Cincinnati OH 45221}
I.~C.~Lai,$^{1}$ S.~Schrenk,$^{1}$
G.~Bonvicini,$^{2}$ D.~Cinabro,$^{2}$ L.~P.~Perera,$^{2}$
G.~J.~Zhou,$^{2}$
G.~Eigen,$^{3}$ E.~Lipeles,$^{3}$
% ...
\end{center}
 
\small
\begin{center}
$^{1}${Virginia Polytechnic Institute and State University,
Blacksburg, Virginia 24061}\\
$^{2}${Wayne State University, Detroit, Michigan 48202}\\
$^{3}${California Institute of Technology, Pasadena, California 91125}\\
% ...
$^{13}${Carleton University, Ottawa, Ontario, Canada K1S 5B6 \\
and the Institute of Particle Physics, Canada}\\
$^{14}${McGill University, Montr\'eal, Qu\'ebec, Canada H3A 2T8 \\
and the Institute of Particle Physics, Canada}\\
% ...
$^{25}${Syracuse University, Syracuse, New York 13244}\\
% ...
\end{center}

\setcounter{footnote}{0}
}
\newpage
Studies of the evolution of a $\KZ$ or $\BZ$ into the respective
anti-particle, a $\KZB$ or $\BZB$,
have guided the form and content
of the Standard Model, and permitted
useful estimates of the 
masses of the charm~\cite{goodetal,glr} 
and top quark~\cite{Albrecht,rosner87} prior to 
their direct observation.
In this Letter, we present
the results of a search for the evolution of the $\DZ$ into
the $\DZB$~\cite{cc}.
Our principal motivation is to observe
new physics outside the Standard Model.

A $\DZ$ can evolve into a $\DZB$ through on-shell intermediate
states, such as $K^+K^-$ with mass, $m_{K^+K^-}\!=\!m_{\DZ}$, or through
off-shell intermediate states, such as those that might be present
due to new physics.  We denote the amplitude through the former (latter)
states by $-iy$ ($x$), in units of $\Gamma_{\DZ}/2$~\cite{ampli}. 
Many predictions for $x$ in the $\DZ\!\to\!\DZB$ amplitude have
been made~\cite{hnncomp}.  The Standard Model contributions are suppressed
to $|x|\approx\tan^2\theta_C\approx5\%$ because
$\DZ$ decay is Cabibbo-favored; the
GIM~\cite{gimktwz} cancellation could further
suppress $|x|$ down to $10^{-6}-10^{-2}$.
Many non-Standard Models
predict $|x| > 1\%$.  Contributions
to $x$ at this level could result from
the presence of new particles with masses as high 
as 100-1000~TeV~\cite{lns,ark}.
Signatures of new physics include
$|x|\!\gg\!|y|$, or charge-parity (CP) violating interference between
$x$ and $y$, or between $x$ and 
a direct decay amplitude.  In order to assess the origin
of a $\DZ\!-\!\DZB$ mixing signal, the effects described by $y$ must be
distinguished from those described by $x$.

The \hbox{wrong-sign} (WS) process, $\DZ\!\to\!K^+\pi^-$, can proceed either
through direct doubly Cabibbo-suppressed decay (DCSD),
or through state-mixing followed by the Cabibbo-favored
decay (CFD), $\DZ\!\to\!\DZB\!\to\!K^+\pi^-$.  
Both processes could contribute
to the time integrated WS rate $R=(f+{\overline f})/2$, 
and the inclusive CP asymmetry 
$A=(f\!-\!{\overline f})/(f\!+\!{\overline f})$, where 
$f=\Gamma(\DZ\!\to\!K^+\pi^-)/
\Gamma(\DZB\!\to\!K^+\pi^-)$, and $\overline{f}$ is defined by
the application of charge-conjugation to $f$.

To disentangle the processes that could contribute to
$\DZ\!\to\!K^+\pi^-$, we study the distribution of WS
final states as a function of the proper decay time $t$
of the $\DZ$.  We describe the proper decay time in units of 
the mean $\DZ$ lifetime, $\tau_{\DZ}=415\pm4\,$fs~\cite{RPP98}.
The differential WS rate 
is~\cite{timev,bsn}
\begin{equation}
r(t)\equiv[R_D + \sqrt{R_D} y^\prime t + 
         \displaystyle{1\over4}(x^{\prime2}\!+\!y^{\prime2})t^2]e^{-t}.
\label{eq:rws}
\end{equation}
The modified mixing amplitudes $x^{\prime}$ and $y^{\prime}$ in
Eqn.~\ref{eq:rws} are given by
$y^\prime \equiv y\cos\delta\!-\!x\sin\delta$,
$x^\prime \equiv x\cos\delta\!+\!y\sin\delta$ and
$R_{M} \equiv \frac{1}{2}\left(x^2+y^2 \right) =
\frac{1}{2}\left(x^{\prime 2}+y^{\prime 2} \right)$
where $\delta$ is a possible strong phase between the DCSD and CFD
amplitudes.
The symbol $R_D$ ($R_M$) represents the DCSD (mixing) 
rate relative to the CFD rate.
There are theoretical arguments 
that $\delta$ is small~\cite{wfsp}, which have
recently been questioned\cite{fnp}.

The influence of each of $x^{\prime}$, $y^{\prime}$, and $R_D$ 
on $r(t)$ in Eqn.~\ref{eq:rws} is distinguishable.
Such behavior is complementary
to the time dependence of the decay rate to CP eigenstates such as
$\DZ\!\to\!K^+K^-$ that is primarily sensitive to $y$, or that of
$\DZ\!\to\!K^+\ell^-\overline{\nu}_{\ell}$ that is sensitive to
$R_{M}$ alone.

We characterize
the violation of CP in state-mixing, direct decay, and the
                interference between those two processes,
                respectively,
                by the real-valued parameters $A_M$, $A_D$, and
                $\phi$,
where, to leading order, both $x^{\prime}$ and $y^{\prime}$ are scaled by
$(1\pm A_M/2)$,
$R_{D}\!\rightarrow\!R_{D}\left(1\pm A_D \right)$,
$\delta\!\rightarrow\!\delta \pm \phi$ in Eqn.~\ref{eq:rws}~\cite{nir99}. 
The plus (minus) sign is used for
an initial $\DZ \ (\DZB)$.

Our data were accumulated between Feb.~1996 and
Feb.~1999 from an integrated luminosity of 9.0~fb$^{-1}$
of $e^+e^-$ collisions at $\sqrt{s}\approx10\,$GeV provided by 
the Cornell Electron Storage Ring (CESR).
The data were taken with the CLEO II multipurpose 
detector~\cite{ctwo}, upgraded in
1995 when a silicon vertex detector (SVX) was installed~\cite{csvx}
and the drift chamber gas was changed 
from argon-ethane to helium-propane.
The upgraded configuration is named CLEO II.V.

We reconstruct candidates for the decay sequence
$D^{*+}\!\to\!\pi^+_s\DZ$, $\DZ\!\to\!K^{\pm}\pi^{\mp}$. 
The charge of the slow
pion ($\pi^+_s$ or $\pi^-_s$) identifies the charm state 
at $t=0$ as either $\DZ$ or $\DZB$. 
We require the $D^{*+}$ momentum,
$p_{D^*}$, to exceed 2.2 GeV, and we require the
$\DZ$ to produce either the final state
$K^+\pi^-$ (WS) or $K^-\pi^+$ (\hbox{right-sign} (RS)). 
The broad features of the
reconstruction are similar to those employed in the recent
CLEO measurement of the $D$ meson lifetimes~\cite{dlife}.

The SVX provides precise measurement of the charged particle 
trajectories, or `tracks,' in three dimensions~\cite{svxres}.
We are thus able to refit the 
$K^+$ and $\pi^-$ tracks with a requirement that they
form a common vertex in three dimensions, and require
that the confidence level (C.L.) of the
refit exceed $0.01\%$.  We use the trajectory of the
$K^+\pi^-$ system and the position of the CESR luminous region
to obtain the $\DZ$ production point.  
We refit the $\pi_s^+$ track with a requirement that
the trajectory intersect
the $\DZ$ production point, and require that the
confidence level of the refit exceed $0.01\%$.
                  
We reconstruct the energy released in the 
$\DZS\!\to\!\pi^+_s\DZ$ decay as $Q\!\equiv\!M^*\!-\!M\!-\!m_\pi$, 
where $M^*$ is the reconstructed mass of the
$\pi_s^+ K^+ \pi^-$ system, $M$ is the reconstructed mass of the
$K^+\pi^-$ system, and $m_\pi$ is the charged pion mass.  
The addition of the $\DZ$ production point to the $\pi_s^+$
trajectory, as well as track-fitting improvements,
yields the resolution $\sigma_Q = 190\pm$6 keV~\cite{thesis}.
The use of helium-propane, in addition to improvements in
track-fitting, yields the resolution
$\sigma_M = 6.4\pm0.1\,$MeV~\cite{thesis}.
These resolutions are better than those 
of earlier studies~\cite{tl,E691,E791,alep}, 
and permit improved suppression of background processes.

Candidates must pass two kinematic requirements
designed to suppress backgrounds from $\DZ\!\to\!\pi^+\pi^-$,
$\DZ\!\to\!K^+K^-$, $\DZ\!\to\,$multibody, and from cross-feed
between WS and RS decays. We evaluate
the mass $M$ for $\DZ\!\to\!K^+\pi^-$ candidates under the
three alternate hypotheses $\DZ\!\to\!\pi^+\pi^-$,
$\DZ\!\to\!K^+K^-$, and $\DZ\!\to\!\pi^+K^-$.  If any one of the
three masses falls within $4\,\sigma$~\cite{masssig}
of the nominal $\DZ$ mass~\cite{RPP98}, 
the $\DZ\!\to\!K^+\pi^-$  candidate is rejected.
A conjugate requirement is made for the RS decays.
The second kinematic requirement rejects 
asymmetric $\DZ$ decays where the pion candidate has low momentum
with the requirement that $\cos\theta^*\!>\!-0.8$ where $\theta^*$ is
the angle of the pion candidate
in the $\DZ$ rest frame with respect to the $\DZ$ boost.
The relative efficiency for the CFD to pass the two
kinematic requirements is $84\%$ and $91\%$, respectively.
 
We require the specific ionization ($dE/dx$) 
measured in the drift chamber for each
charged track agree to within 3$\sigma$ of the
expected value; this is a loose criterion, and we vary the $dE/dx$
requirement for systematic studies.
\begin{figure}[htb]
\begin{center}
\epsfig{figure=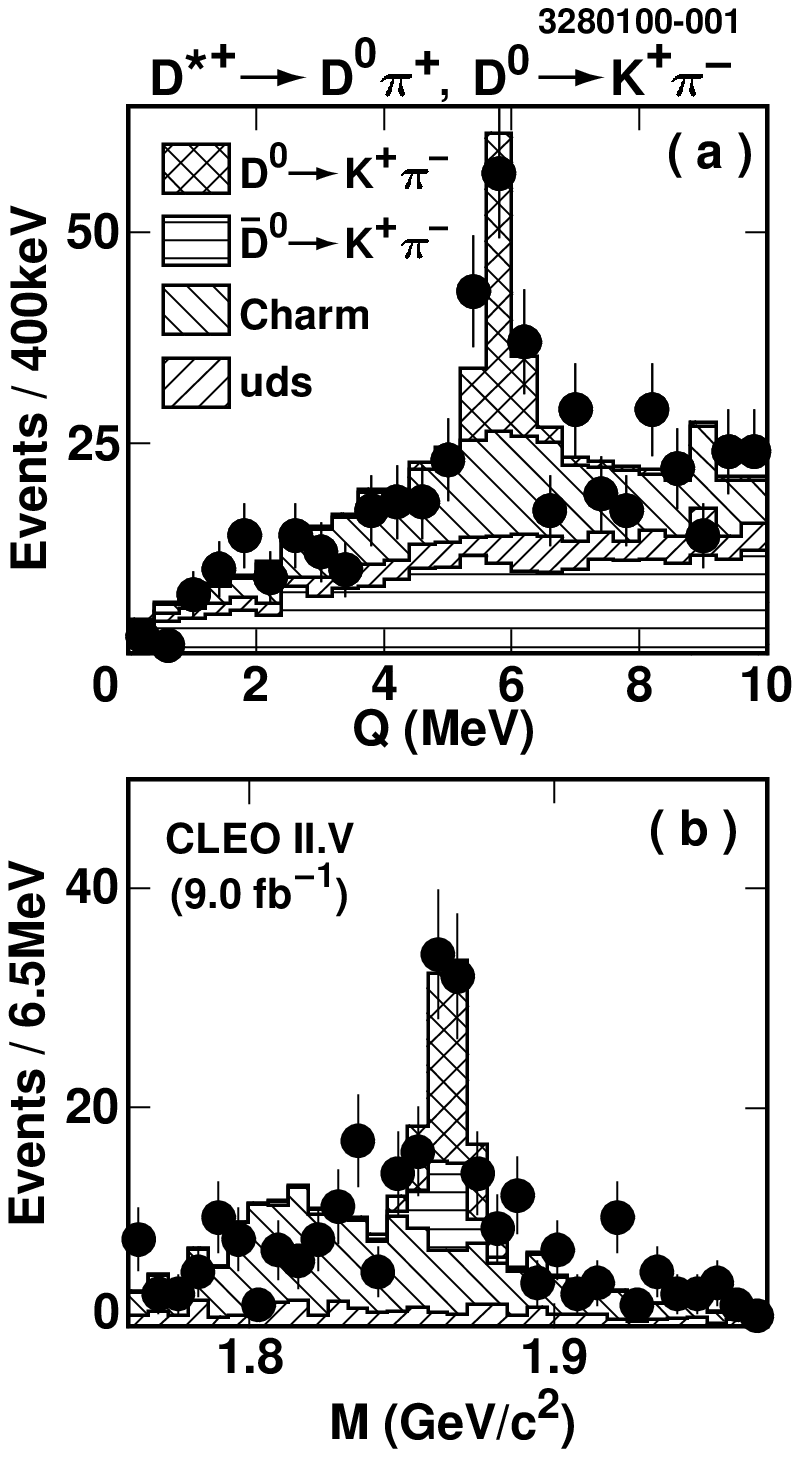,width=4in}
\end{center}
\caption[Signal for the WS Process $\DZ\!\to\!K^+\pi^-$]
{Signal for the WS process $\DZ\!\to\!K^+\pi^-$.
The data are the full circles with error bars, the projection of the
fit for the signal
is cross-hatched, and the projections of the 
fit for the backgrounds from charm and light quark
production are single-hatched.
For part a), $M$ is within $2\sigma$~\cite{masssig} 
of the CFD value, and
for b), $Q$ is within $2\sigma$~\cite{masssig} 
of the CFD value.}
\label{fig:qmws}
\end{figure}

We reconstruct $t$ using only the vertical ($y$) component of the
flight distance of the $\DZ$ candidate.  
This reconstruction is effective because
the vertical extent of the $e^+e^-$ luminous region has
$\sigma_y=7\,\mu m$\cite{Cinabro:1997ar}.
The resolution on the
$\DZ$ decay point $(x_v,y_v,z_v)$ is
typically $40\,\mu$m in each dimension.   We measure the
centroid of the luminous region $(x_b,y_b,z_b)$ with
hadronic events in blocks of data with integrated
luminosities of several pb$^{-1}$, and an error on
$y_b$ that is less than $5 \mu m$.
We reconstruct $t$ as $t=M/p_y \times (y_v-y_b)/(c\tau_{\DZ})$
where $p_y$ is the $y$ component of the total momentum of the
$K^+\pi^-$ system.  The error in $t$, $\sigma_t$, 
is typically $0.4$ (in $\DZ$ lifetimes),
although when the $\DZ$ direction is near the horizontal plane, $\sigma_t$
can be large; we reject $12\%$ of the CFD by requiring $\sigma_t < 3/2$.
Studies of the plentiful RS sample allow us to determine our
resolution function~\cite{dlife}, and show that biases
are negligible for the WS results.

Our signal for the WS process $\DZ\!\to\!K^+\pi^-$ is shown
in Fig.~\ref{fig:qmws}.  We determine the background levels by performing
a fit to the two-dimensional region of $0\!<\!Q\!<\!10\,$MeV versus
$1.76\!<\!M\!<\!1.97\,$GeV that has an area 135 times larger
than our signal region.  Event samples generated by the Monte Carlo (MC)
method and fully simulated in our detector\cite{mcleo} 
corresponding to $90\,$fb$^{-1}$
of integrated luminosity are used to estimate the background shapes
in the $Q\!-\!M$ plane.  The normalizations of the background
components with distinct distributions in the $Q\!-\!M$ plane
are allowed to vary in the fit to the data.  
The background distributions
and normalizations in the $\DZ$ and $\DZB$ samples
are consistent and constrained to be identical. 
We describe the signal shape in the $Q\!-\!M$ plane 
with the RS data that is within
$4\,\sigma$ \cite{masssig} of the nominal CFD value. 
The results of the fit are displayed in Fig.~\ref{fig:qmws} 
and summarized in Table~I.

\begin{center}
\begin{minipage}[!]{.9\textwidth}
TABLE~I. Fitted event yields in a region of $2\,\sigma$~\cite{masssig} 
centered on the CFD $Q$ and $M$ values. The
total number of candidates is 82. The estimated background is $37.2\pm1.8$.
The bottom row describes the normalization sample.
\vspace{-1mm}
\begin{center}
\begin{tabular}{ll}
\hphantom{aa}\textbf{Component} &
\hphantom{aa}\textbf{Yield} \\ \hline \hline
\hbox to0pt{\raise 0.75ex\hbox to0pt{\vphantom{X}}}
$\DZ\!\to\!K^+\pi^-$ (WS) & \hphantom{a} $44.8^{+9.7}_{-8.7}$  \\
\hbox to0pt{\raise 0.75ex\hbox to0pt{\vphantom{X}}}
random $\pi^{+}_s\!$, $\DZB\!\to\!K^+\pi^-$   & \hphantom{a} $16.0\pm1.6$  \\
\hbox to0pt{\raise 0.00ex\hbox to0pt{\vphantom{X}}}
{$e^+e^-\!\rightarrow\!c\overline{c}$}~~bkgd.& \hphantom{a} {$17.6\pm0.8$}  \\
\hbox to0pt{\raise 0.00ex\hbox to0pt{\vphantom{X}}}
{$e^+e^-\!\rightarrow\!u\overline{u},d\overline{d},s\overline{s}$}~~bkgd. &
\hphantom{a} {$\hphantom{1}3.6\pm0.4$} \\ \hline \hline
\hbox{\raise 0.75ex\hbox{\vphantom{X}}}
$\DZB\!\to\!K^+\pi^-$ (RS)\hspace{0.5cm} & $13527\pm116$ \\ \hline
 \hline \vspace{1mm}
\end{tabular}
\end{center}
\end{minipage}
\end{center}

The proper decay time distribution is shown
in Fig.~2 for WS candidates that are within
2$\sigma$~\cite{masssig} 
of the CFD signal value in the $Q\!-\!M$ plane.
We performed maximum-likelihood
fits in bins that are 1/20 of the $\DZ$ lifetime.
The background levels are constrained to
the levels determined in the fit to the $Q\!-\!M$ plane. 
We use the resolution function in $t$ to describe
the $e^+e^-\!\rightarrow\!
u\overline{u},d\overline{d},s\overline{s}$ backgrounds,
and an exponential, folded with the resolution function, to describe
the $e^+e^-\!\rightarrow\!c\overline{c}$ backgrounds.  
The distribution in $t$ of the RS data is used to represent the
random $\pi^{+}_s\!$,~$\DZB\!\to\!K^+\pi^-$ background~\cite{bkgdlife}.
The WS signal is described by Eqn.~\ref{eq:rws}, either
modified to describe all three forms of CP violation (Fit A),
without modification to describe mixing alone (Fit B), or with the
mixing parameters constrained to be zero (Fit C).  The effect of
our resolution is always included.

The reliability of our fit depends
upon the simulation of the decay time distribution of
the background in the signal region. 
A comparison of the proper time for the data and MC
samples for several sideband regions yields a
$\chi^2=4.4$ for 8 degrees of freedom and 
supports the accuracy of the background simulation~\cite{thesis}.

Our principal results concerning mixing are determined from Fit A.
The one-dimensional, 95\% confidence intervals for $x^{\prime}$,
$y^{\prime}$ and $R_D$, determined by 
an increase in negative log likelihood ($-\ln{\cal L}$) of 1.92,
are given in Table~II. 
The fits are consistent with an absence of both mixing and
CP violation. The small change in
likelihood when mixing and CP violation are allowed could be
a statistical fluctuation, or an emerging signal.

\begin{center}
\begin{minipage}[!]{.9\textwidth}
TABLE~II. Results of the fits to the $\DZ\!\to\!K^+\pi^-$ decay time
distribution. Fit A allows
both $\DZ\!-\!\DZB$ mixing and CP violation. 
In Fit B, we constrain $A_M$, $A_D$, and $\phi$ to zero.
In Fit C, we constrain $x^\prime$ and $y^\prime$ to zero,
so $R\equiv R_D$ and $A\equiv A_D$.
The incremental change in $-\ln{\cal L}$
for Fit B (Fit C) with respect
to the Fit A (Fit B) is 0.07 (1.57).
\begin{center}
\begin{tabular}{ccc}
\textbf{Parameter} & \textbf{Best Fit} & \textbf{95\% C.L.}\\ \hline \hline
Fit A \hfill \hphantom{a} & \multicolumn{2}{c}{Most General Fit} \\
\vphantom{$A^{A^A}$}$R_D$      & $(0.48\pm0.12\pm0.04)\%$ & 
$0.24\%<R_D<0.71\%$\\
$y^\prime$ & $(-2.5^{+1.4}_{-1.6}\pm0.3)\%$ & $-5.8\%<y^\prime<1.0\%$ \\ 
$x^\prime$ & $(0\pm1.5\pm0.2)\%$             & $|x^\prime|<2.9\%$ \\
$(1/2)x^{\prime2}$ &     & $<0.041\%$ \\ \hline 
 & \multicolumn{2}{c}{CP violating parameters} \\
$A_M$ & $0.23^{+0.63}_{-0.80}\pm 0.01$ & No Limit \\
$A_D$ & $-0.01^{+0.16}_{-0.17}\pm 0.01$ & $-0.36<A_D<0.30$\\
$\sin \phi$ & $0.00 \pm 0.60\pm 0.01$ & No Limit \\ \hline \hline
Fit B \hfill \hphantom{a} & \multicolumn{2}{c}{CP conserving fit}  \\
\vphantom{$A^{A^A}$}$R_D$      & $(0.47^{+0.11}_{-0.12}\pm0.04)\%$ & 
$0.24\%<R_D<0.69\%$\\
$y^\prime$ & $(-2.3^{+1.3}_{-1.4}\pm0.3)\%$ & $-5.2\%<y^\prime<0.2\%$ \\ 
$x^\prime$ & $(0\pm1.5\pm0.2)\%$             & $|x^\prime|<2.8\%$ \\
$(1/2)x^{\prime2}$ &     & $<0.038\%$ \\ \hline \hline
Fit C \hfill \hphantom{a} & \multicolumn{2}{c}{No-mixing fit \hfill
$R \equiv R_D$ , $A \equiv A_D$} \\
$R$ & \multicolumn{2}{c}{$(0.332^{+0.063}_{-0.065}\pm0.040)\%$}  \\
$A$ & $-0.02^{+0.19}_{-0.20}\pm 0.01$ & $-0.43<A<0.34$ \\ \hline
\hbox to0pt{\raise 0.75ex\hbox to0pt{\vphantom{X}}}
$R/\tan^4\theta_C$ & \multicolumn{2}{c}{$(1.24^{+0.23}_{-0.24}\pm0.15)$}  \\
${\cal B}(\DZ\!\to\!K^+\pi^-)$ &
\multicolumn{2}{c}{$(1.28^{+0.24}_{-0.25}\pm0.15\pm0.03)\!\times\!10^{-4}$} \\ \hline \hline
\end{tabular}
\end{center}
\end{minipage}
\end{center}

We make contours in the two-dimensional
plane of $y^{\prime}$ versus $x^{\prime}$ that
contain the true value of $x^{\prime}$ and
$y^{\prime}$ with 95\% confidence, for Fit A and Fit B.
The contour is where $-\ln{\cal L}$ increases
from the best fit value by 3.0.
All fit variables other than $x^{\prime}$ and $y^{\prime}$
are allowed to vary to best fit
values at each point on the contour.
The interior of the contour is the tightly
cross-hatched region near the origin of Fig.~\ref{fig:xyl}.
The limits are not substantially degraded
when the most general CP violation is allowed,
in part because our acceptance, unlike that of earlier
studies~\cite{E691,E791}, is uniform as a function
of $\DZ$ \ decay time.
\begin{figure}[htb]
\begin{center}
\epsfig{figure=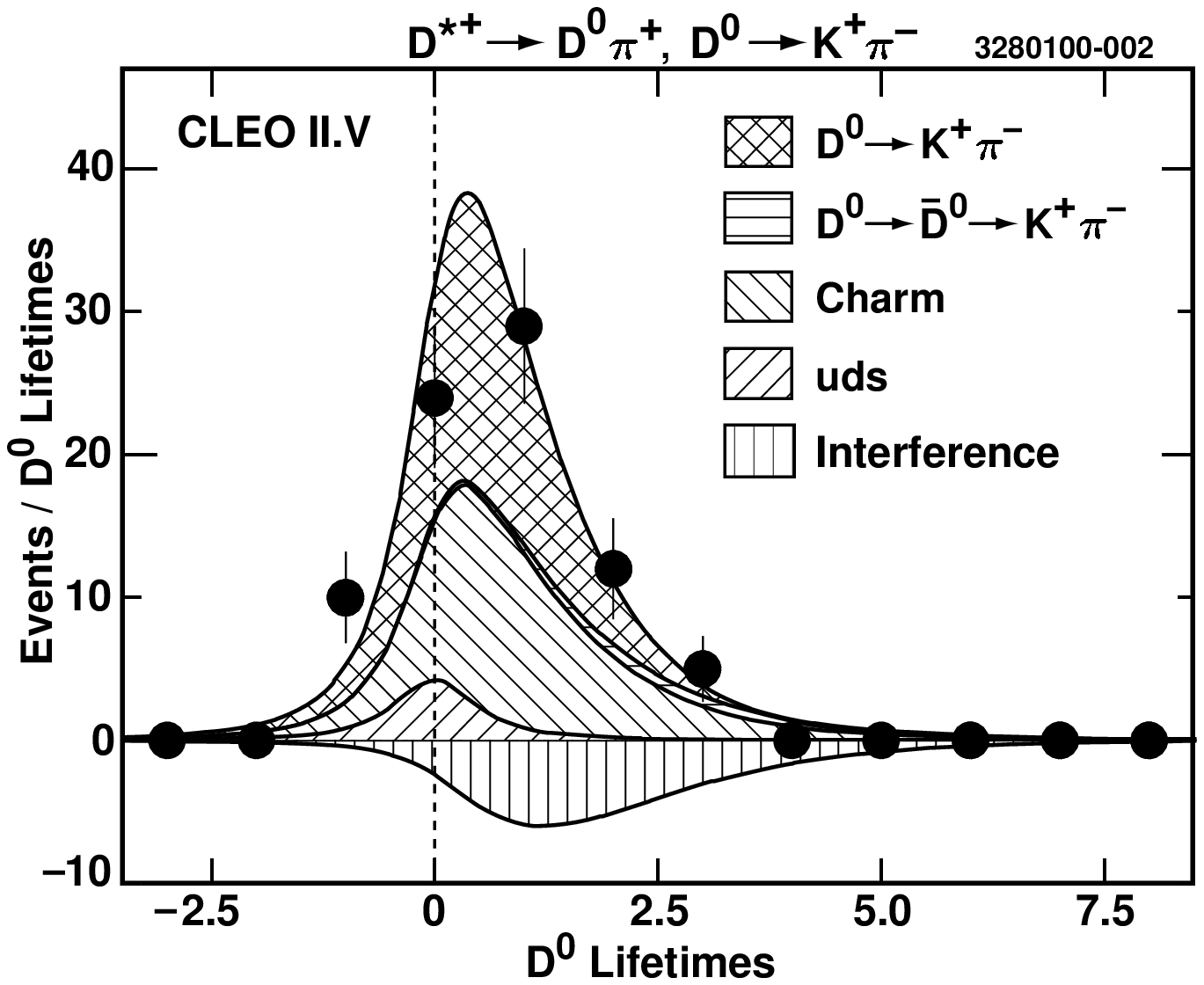,width=5in}
\end{center}
\caption[Distribution in $t$ for $\DZ\!\to\!K^+\pi^-$.]
{Distribution in proper decay time for
$\DZ\!\to\!K^+\pi^-$ candidates, and the best fit of type A,
described in Table~II.
The data are shown as the full circles with error bars.
The cross-hatched region is the sum of the fit contribution
from the direct $\DZ\!\to\!K^+\pi^-$ decay and the fit contribution
from the destructive interference with mixing, 
which is shown in the region with single, vertical hatching.
The fit 
contributions from backgrounds charm and light quark production are shown
in single, diagonal hatching.}
\label{fig:lws}
\end{figure}

Many classes of systematic error cancel
due to the similarity of the events that comprise the numerators and 
denominators of $R$ and $A$. The dominant systematic errors stem from
potential misunderstanding of the shapes and acceptances for our
backgrounds.  We vary the selection criteria to estimate these
systematic errors from the data. The level and composition of the
backgrounds are sensitive to the requirements on 
momentum magnitude and direction, and $dE/dx$
of the charged particle trajectories and contribute $\pm 0.018\%$, $\pm
0.018\%$ and $\pm 0.026\%$, respectively, to the systematic error in $R_D$.
We also include the 
statistical uncertainty on the MC determination of the
proper time for the $e^+e^-\!\rightarrow\!q\overline{q}$ 
backgrounds~\cite{bkgdlife} in the systematic error. 
We assess a total systematic error on $R_D$, $x^\prime$ and $y^\prime$ of 
$\pm0.040\%$, $\pm0.2\%$ and $\pm0.3\%$, respectively.
A study of detector-induced and event-reconstruction-induced asymmetries in CFD
limits the relative systematic error on $A$ to $<\!1\%$.

If we assume that $\delta$ is small, 
then $x^\prime\!\approx\!x$
and we can indicate the impact of our work in limiting predictions
of $\DZ\!-\!\DZB$ mixing from extensions to the Standard Model.
The 95\% C.L. interval for $x$ from Fit A has some inconsistency with
eighteen of the predictions tabulated in Ref.~\cite{hnncomp}.
A new model~\cite{brhlik} invokes SUSY to account for the value of
$\epsilon^\prime/\epsilon$~\cite{epsp}
and estimates $\DZ\!-\!\DZB$ mixing with $x=0.6\%$, somewhat
below our sensitivity. Another analysis~\cite{bn}
notes that SUSY could induce $A_D\!\sim\!30\,$\%, just below
our sensitivity.

In conclusion, our data 
are consistent with no $\DZ\!-\!\DZB$ mixing. 
We limit the mixing amplitudes, $x^{\prime}$ and $y^\prime$, to be
$(1/2)x^{\prime2}<0.041\%$ and
$-5.8\%\!<\!y^\prime\!<\!1.0\%$
at the 95\% C.L., without assumptions
concerning CP violating parameters.
We have observed $44.8^{+9.7}_{-8.7}$ candidates for the decay
$\DZ\!\to\!K^+\pi^-$ corresponding to 
$R=(0.332^{+0.063}_{-0.065}\pm0.040)\%$.
We observe no evidence for CP violation.
These results are a substantial advance in sensitivity to the 
phenomena that contribute to the
wrong-sign process $\DZ\!\to\!K^+\pi^-$.
\begin{figure}[htb]
\vspace{-4cm}
\begin{center}
\epsfig{figure=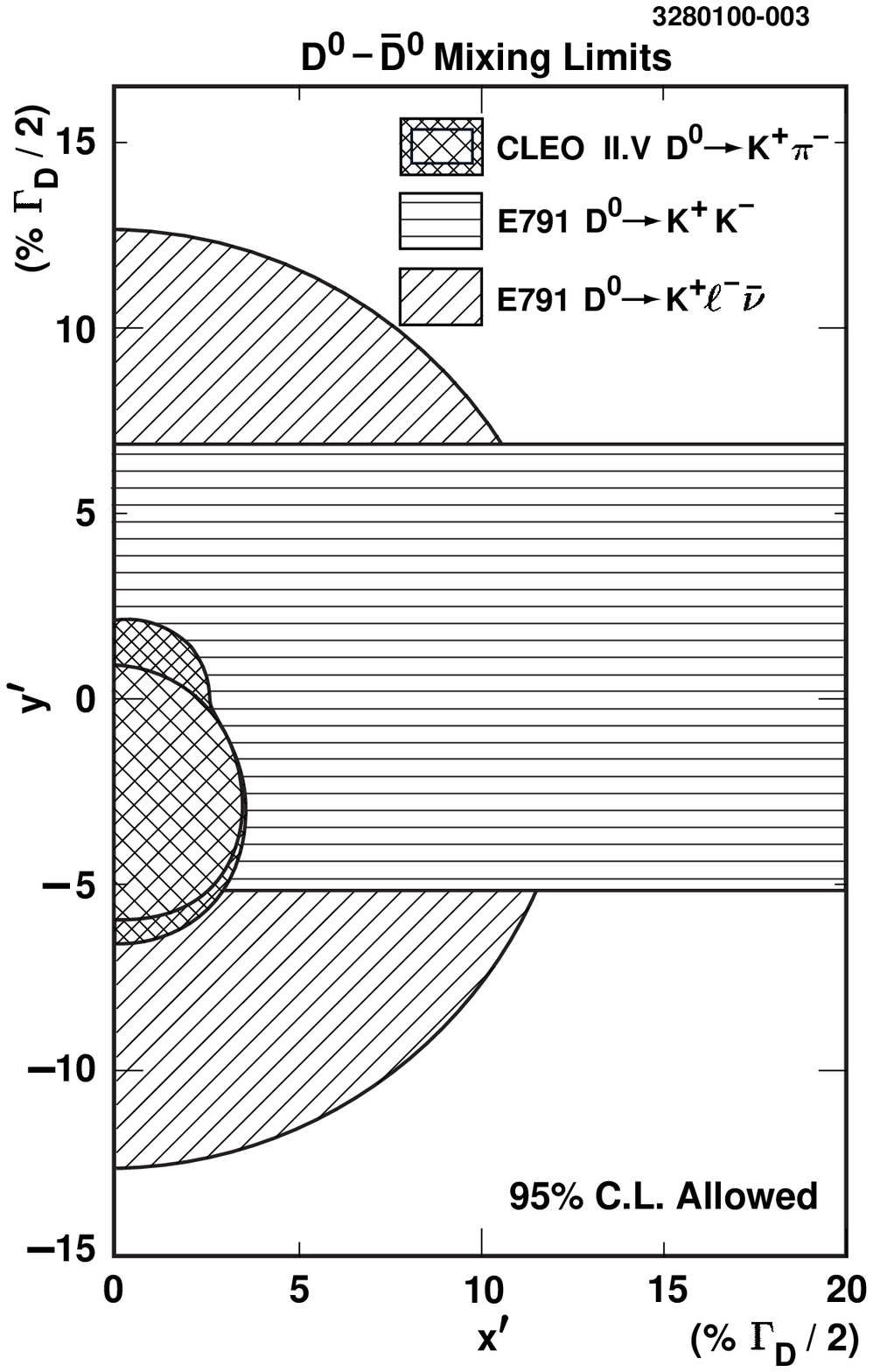,width=5in}
\end{center}
\caption[Allowed regions in the $y^\prime$ vs. $x^\prime$ Plane]
{Allowed regions, at 95\% C.L., 
 in the $y^\prime$ vs. $x^\prime$ plane.  
The entire kidney shaped region, filled with tight cross-hatching,
is allowed under Fit A of Table II, while Fit B, in which CP violation
is assumed, allows the smaller region, which is overlayed and filled
with looser cross-hatching.
The allowed regions from studies comparable
to Fit A, using 
$\DZ\!\to\!K^+K^-$~\cite{E791CP}, for which we assume $\delta\!=\!0$,
and $\DZ\!\to\!K^+\ell^-\overline{\nu}_{\ell}$~\cite{E791L}, 
are shown as single-hatched regions.  The Bayesian approach is
used\cite{RPP98}.}
\label{fig:xyl}
\end{figure}

We gratefully acknowledge the effort of the CESR staff in providing us
with
excellent luminosity and running conditions.
We wish to acknowledge and thank the technical staff who contributed
to
the success of the CLEO~II.V detector upgrade, including J.~Cherwinka
and
J.~Dobbins (Cornell); M.~O'Neill (CRPP); M.~Haney (Illinois);
M.~Studer and
B.~Wells (OSU); K.~Arndt, D.~Hale, and S.~Kyre (UCSB).
We thank Y.~Nir for his help in the development of our description
of CP violating effects.
This work was supported by
the National Science Foundation,
the U.S. Department of Energy,
Research Corporation,
the Natural Sciences and Engineering Research Council of Canada,
the A.P. Sloan Foundation, the Swiss National Science Foundation,
and the Alexander von Humboldt Stiftung.


\begin{thebibliography}{99}
\bibitem{cc}
The processes obtained by charge conjugation to initial through final
states are implied, except in specific instances when CP
violation is considered.
\bibitem{svxres} Impact parameter resolutions as functions of
momentum $p$ (GeV/$c$) are
measured from data to be
$\sigma_{xy} = 19 \oplus 39 / (p\sin^{3/2}\theta)$\
$\mu $m and
$\sigma_{rz} = 42 \oplus 45 / p$\ $\mu $m
at $\theta = 90^\circ$, where $\theta$ is the angle between the charged
particle trajectory and the colliding beam axis.
\bibitem{masssig} 
The standard deviation $\sigma$ for the quantities $Q$ and $M$ is
computed from the trajectory reconstruction covariance matrix of the 
daughters of each $\DZS$ candidate.
We reject poorly constructed candidates where $\sigma_Q$ ($\sigma_M$)
exceeds 300 keV (10 MeV).
\bibitem{bkgdlife} 
The mean decay time of the $e^+e^-\to u{\overline u},d{\overline
d},s{\overline s}$ and $c{\overline c}$ background is determined from the MC
sample to be $2.6\pm 6.6\,$fs and $408.6\pm 27.3\,$fs, respectively.
From the RS data sample, 
we determine $\tau_{\DZ}=400.7\pm3.9\,$fs without assessment of the systematic
uncertainty.
This result does not supersede~\cite{dlife}.
From the MC sample, we determine, $\tau_{\DZ}=416.0\pm1.6\,$fs, consistent
with the input value of $415\,$fs.
\end{thebibliography}
\end{document}